\documentclass[11pt,letterpaper]{article}
\usepackage{graphicx}
\usepackage{bm}
\usepackage{amssymb}
\usepackage{color}
\usepackage{amsmath}
\usepackage[numbers]{natbib}
\usepackage{enumerate}
\usepackage{amstext}
\usepackage{footmisc}
\usepackage{latexsym}
\usepackage[usenames,dvipsnames]{xcolor}
\usepackage[colorlinks=true,citecolor=Cerulean,linkcolor=RubineRed,urlcolor=Cerulean]{hyperref}
\usepackage{multibib}
\usepackage{ fixfoot}
\usepackage{fancyhdr}
\newcommand{\pap}[1]{\left( #1 \right)}

\DeclareFixedFootnote{\rep}{Electronic address: \href{mailto:fernandojavier.gomez@iff.csic.es}{fernandojavier.gomez@iff.csic.es}}

\newcommand{\beq}{\begin{equation}}
\newcommand{\eeq}{\end{equation}}
\newcommand{\beqa}{\begin{eqnarray}}
\newcommand{\eeqa}{\end{eqnarray}}

\begin{document}
\title{{\bf Vulnerability of Quantum Information Systems to Collective Manipulation}}
\author{Fernando J. G\'omez-Ruiz\href{https://orcid.org/0000-0002-1855-0671}{\includegraphics[scale=0.45]{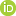}}\,$^{1}$, Ferney J. Rodr\'iguez\href{https://orcid.org/0000-0001-5383-4218}{\includegraphics[scale=0.45]{orcid}}\,$^{2}$, 
Luis Quiroga\href{https://orcid.org/0000-0003-2235-3344}{\includegraphics[scale=0.45]{orcid}}\,$^{2}$, and Neil F. Johnson\href{https://orcid.org/0000-0002-3224-3213}{\includegraphics[scale=0.45]{orcid}}\,$^{3,\dagger}$}
\date{}
\maketitle
\vspace{-1cm}
\begin{center}
$^{1}${\it Departamento de F\'isica Te\'orica, At\'omica y \'Optica, Universidad de Valladolid, 47011 Valladolid, Spain}\\
$^{2}${\it Departamento de F{\'i}sica, Universidad de los Andes, A.A. 4976, Bogot\'a D. C., Colombia}\\
$^{3}${\it Physics Department,  George Washington University, Washington D.C. 20052, U.S.A.}\\
$^{\dagger}${\it Corresponding Author:} \href{mailto:neiljohnson@gwu.edu}{neiljohnson@gwu.edu}
\end{center}
\begin{abstract}
The highly specialist terms `quantum computing' and `quantum information', together with the broader term `quantum technologies', now appear regularly in the mainstream media. While this is undoubtedly highly exciting for physicists and investors alike, a key question for society concerns such systems' vulnerabilities -- and in particular, their vulnerability to collective manipulation. Here we present and discuss a new form of vulnerability in such systems, that we have identified based on detailed many-body quantum mechanical calculations. The impact of this new vulnerability is that groups of adversaries can maximally disrupt these systems' global quantum state which will then jeopardize their quantum functionality. It will be almost impossible to detect these attacks since they do not change the Hamiltonian and the purity remains the same; they do not entail any real-time communication between the attackers; and they can last less than a second. We also argue that there can be an implicit amplification of such attacks because of the statistical character of modern non-state actor groups. A countermeasure could be to embed future quantum technologies within redundant classical networks. We purposely structure the discussion in this chapter so that the first sections are self-contained and can be read by non-specialists. 
\\
{\bf DOI:} \href{https://www.intechopen.com/online-first/1174060}{10.5772/intechopen.1004935}
\end{abstract}

\section{Introduction} 
Technology is rapidly moving toward an era that will fully embrace the true ``spookiness" (e.g. action-at-a-distance) of quantum mechanics, where quantum mechanical information processing systems will offer unique advantages over current classical counterparts~\cite{Ph_World,Ekert,Lloyd,Benjamin,Ekert2,Zeilinger2,lidar2,Zurek1,Zurek2,Zeilinger1,Kaku,arxiv}. Enormous investments have been made recently in quantum information technologies: in the United States with the NSF  `Quantum Leap' initiative~\cite{NSF}, in Europe~\cite{EU_Fund}  and in China~\cite{China_Fund}. Among other global-scale applications, the idea of a quantum Internet is gaining significant traction~\cite{pirandola,Davide,Wehner,Pan1,Pan2,Cartwright,arxiv}. A key desirable feature of such a quantum Internet is that it provides enhanced security and novel functionalities not present in the existing classical Internet infrastructure. Human nature, however, will likely remain unchanged \cite{arxiv}. Just as human ingenuity can be used for good to drive quantum technology -- which is the overriding narrative of most funding agencies~\cite{NSF,arxiv} -- it could also be used for bad~\cite{Gill1,Gill2,Neil2,us2,us3,us4,us5,Neil3}. An obvious yet largely unanswered question then arises, of what might go wrong at scale in such a quantum system in terms of bad-actor manipulation?\\
\begin{figure}[t!]
\begin{center}
\includegraphics[width=1.0\linewidth]{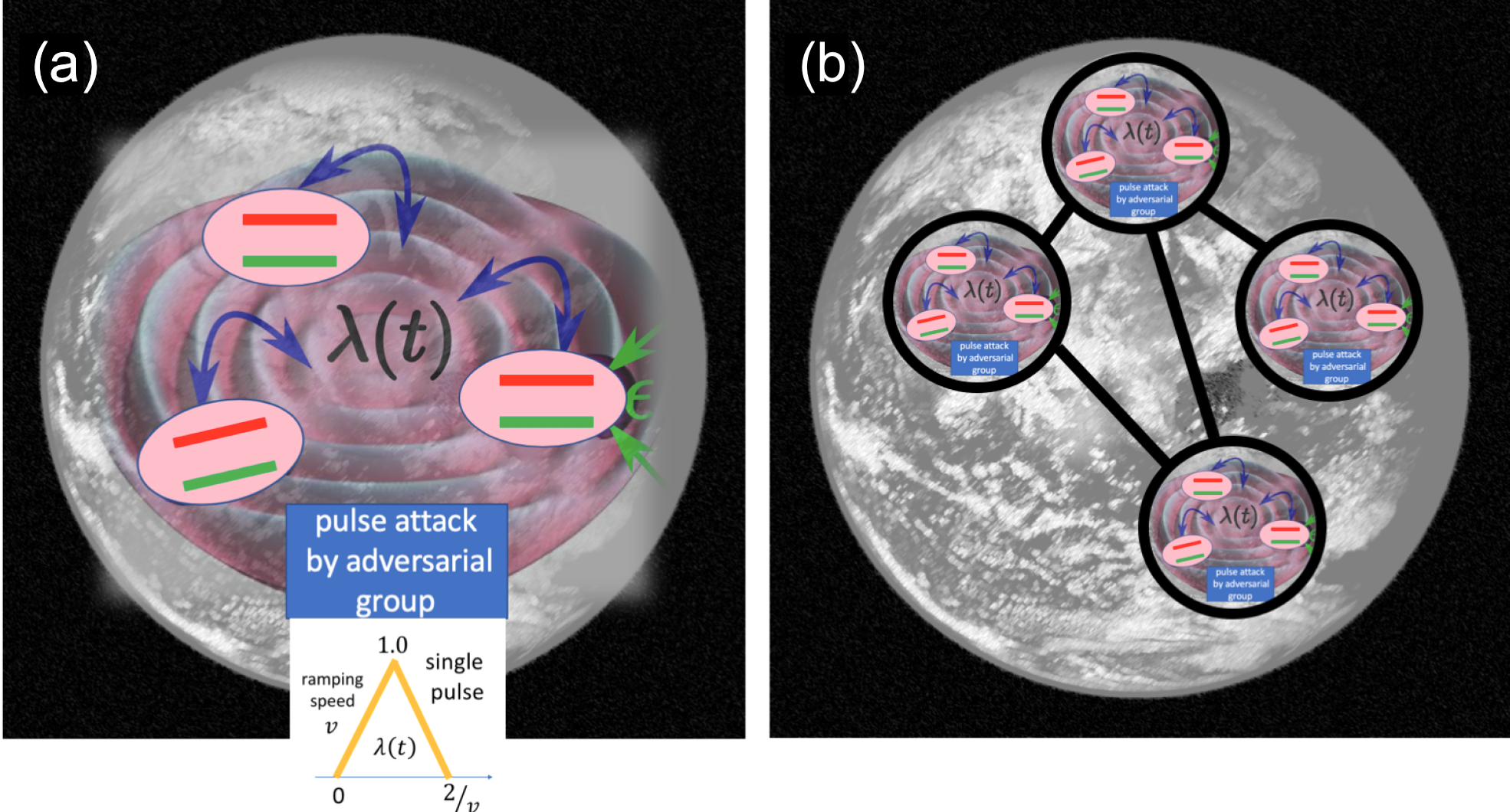}
\end{center}
	\caption{\label{fig1} {\bf Future quantum technologies.} {\bf (a)} Ultimate quantum technology limit of an extended geographical space covered by a quantum cloud in which the global quantum state is coherent~\cite{arxiv}. This could be a spatially extended cavity of bosonic modes containing an arbitrary number $N$ of qubits (i.e. two-level systems). A group of adversaries attack by applying a pulse interaction $\lambda(t)$ (see Eq.~\eqref{hdic}) between the qubits and a bosonic mode at speed $\upsilon$, with a duration $2/\upsilon$. Our conclusions are insensitive to the precise shape of the pulse, but we show a triangular pulse here for concreteness. {\bf (b)} An intermediate, simpler version that could be built sooner than {\bf (a)} because of lower technological demand. It features smaller versions of the $N=3$ qubit-cavity system in {\bf (a)}, which are then interconnected through separate quantum or classical communication channels.}
\end{figure}

Based on the recent successes of building and manipulating matter in the form of nanostructure systems, and light in the form of photon fields as carriers of information, it seems likely that a future quantum information processor or quantum Internet will comprise a matter-light system. In other words, it could easily end up containing $N$ qubits (two-level quantum systems) made of matter that sit in some photonic environment (i.e. light) and hence bosonic field as in Fig.~\ref{fig1}(a).\\ 
\\
We hence focus our discussion in this chapter around this generic system setup in Fig.~\ref{fig1}(a). As we will go on to explain in this chapter, our interest lies in seeing what the effect would be if $N$ adversaries -- who each choose one of the $N$ qubits to attack  -- simultaneously manipulate the coupling $\lambda$ between their chosen qubit and one of the bosonic modes by ramping $\lambda$ up and down as in Fig.~\ref{fig1}(a). The resulting pulse profile $\lambda(t)$ applied simultaneously between all $N$ qubits and the bosonic mode  is what we mean by collective manipulation of the quantum system by a hostile group, and the extent of the damage (i.e. change) that they cause to the system's initial overall quantum state will act as an indicator of the system's vulnerability to that type of collective attack.
To provide some practical background, there is a long history of theoretical and empirical research exploring the properties of individual quantum nanoscale systems as potential qubits, e.g. a single superconductor nanostructure, or a single trapped ion, or a single semiconductor quantum dot~\cite{qudots}. The semiconductor quantum dot example, though perhaps less well-known, is essentially identical to the others in that it is a nanoscale system which has (at least) two well-defined discrete energy levels, each  with their own wavefunctions, and hence has been termed an `artificial atom' in the literature. The focus has been on the properties of the ground and excited states and how the quantum information in this system could be manipulated by external magnetic or electric fields, or by some particular photonic input~\cite{LaurentPRX}. The goal of most such studies has been to explore how the system can be switched from one eigenstate to another, as well as creating quantum superpositions of such states. There is also a lot of work on pairs of qubits (e.g. two superconductor nanostructures, two trapped ions, or two semiconductor quantum dots). This is  because such a pair ($N=2$) represents the smallest example of a system where quantum entanglement can be explored in spatially separate systems. The goal in that case is to understand better how to exploit the `spooky action at a distance' property of the quantum world, in order to store and transfer quantum information and develop quantum gates as part of some larger scale quantum information processor, quantum computer or  quantum network. These studies can be challenging both experimentally and theoretically, hence the question of what might happen with a larger number of qubits (e.g. $N=3,4\dots$ as in Fig.~\ref{fig1}(a)) has received far less attention. Yet recent research~\cite{Gomez_Ruiz_2023} inspired by reports of novel quantum effects in photosynthesis, has shown that new quantum phenomena can arise in systems with $N=3,4\dots$ qubits that do not occur in systems of $N=1$ or $2$ qubits. Given that exotic quantum many-body states like superconductivity are known to exist for $N\rightarrow \infty$, this suggests that new types of quantum information phenomena can also arise for $N=3,4,\dots$ qubits that do not occur for $N=1$ or $2$ qubits.\\
\\
This helps motivate the work in this present chapter, i.e.  to analyze the potential vulnerabilities of systems with $N=3,4,\dots$ etc. qubits in a bosonic environment. Any future quantum information processor or computer will of course contain $N\geq 3$ qubits in total, hence the importance of understanding such a system. In short, one might expect a quantum version of the phrase `two's company but three is a crowd' and hence the vulnerabilities of a (quantum) crowd of qubits in a bosonic mode (i.e. $N\geq 3$) will be different from those of an individual qubit or a pair of qubits in a bosonic mode.\\
\\
Heightening this need to understand  $N\geq 3$ qubit-plus-bosonic-mode vulnerabilities from the quantum hardware side, is the fact that the study of attacks on quantum information systems has tended to focus around scenarios involving the exchange of information (qubit states) between two entities (so-called Alice and Bob). The third entity is simply a single adversary (Eve, the eavesdropper). While such Alice-Bob and Eve thought experiments are essential for understanding what happens to one or two qubits and their information in terms of cryptography and secrecy, there is therefore a gap in understanding the vulnerabilities of systems containing  $N=3$ or more qubits. 
Finally, since the Hilbert space of quantum states explodes in size as $N$ increases, as does  its internal complexity, the properties of an entangled state for $N=3,4\dots$ qubits-plus-bosonic mode can have entirely new opportunities for storing and processing quantum information. Hence the $N=3,4\dots$ qubits-plus-bosonic mode quantum state will likely have entirely new vulnerabilities to attack by adversaries, as we indeed show in this chapter.\\
\\
In the social science field of conflict studies, this need to understand the effects of groups of adversaries is already well recognized, since this is how most insurgent and terrorist attacks occur~\cite{Robb,Gill1,Neil2,us2,us3,us4,us5} -- with the added modern twist (which also applies to quantum information attacks in Fig.~\ref{fig1}(a)) that they need not all be in the same geographical location since they just need a means of synchronizing their actions (e.g. through some common clock) \cite{Gill1,Neil2,arxiv}. In other words, they all initiate the same manipulation $\lambda(t)$ between their chosen qubit and the bosonic mode in Fig.~\ref{fig1}(a) starting at exactly the same time.\\
\\
In summary, this need to improve the understanding of vulnerabilities in $N\geq 3$ qubit-bosonic-mode quantum information systems explains the need for this chapter~\cite{arxiv}. In particular, we will consider the following urgent question: {\em Can new types of vulnerabilities arise in future matter-light quantum technologies which will undoubtedly have  $N\geq 3$ qubits}?
We will show that the answer is a definitive `yes': specifically, we will show that an entirely new form of threat arises by which 
a group of quantum-enabled adversaries can maximally disrupt the global quantum state of systems such as Fig.~\ref{fig1}(a) containing $N\geq 3$ qubits,  in a way that is practically impossible to detect -- and which may even get amplified by the way that humans naturally cluster into adversarial groups \cite{Spagat, Neil2,us3,us4,us5}. We offer a possible countermeasure but stress that there may be entire classes of such threats for which, as yet, there is no underlying scientific theory or understanding. Our analysis leverages the fact that whatever the future quantum technology, the necessary intercommunication across geographical distances will likely rely on electromagnetic waves -- and hence a bosonic field of photons. This reinforces our reason for basing our analysis around a model of a generic, global quantum system involving a bosonic field~\cite{Acevedo2014PRL}.\\
\\
We note that although we ourselves are condensed matter physicists with a background in semiconductors, everything that we discuss concerning qubits in this chapter is general. The individual qubits could be of any form, as long as they each have two identifiable eigenstates and that transitions can be made between them by some external driving field. We expect that some of this book's audience will have more of a quantum computing background (and will more naturally consider qubits more from the perspective of superconducting nanostructures or trapped ions). But in principle the two states per qubit could be anything -- including exotic topological states of some kind. As long as the system can be mapped to the same form as the Hamiltonian in Eq.~\eqref{hdic}, which many can if the appropriate approximations are made, then all our results will hold and our conclusions will be unchanged.\\ 
\\
We also note that some of the content of this chapter appears in earlier working papers on the arXiv preprint server (e.g. Ref.~\cite{arxiv} and to a far lesser extent Ref.~\cite{arxiv2}). However the current chapter is not published anywhere else: pieces were posted on arXiv simply to gather feedback from the quantum information and quantum computing communities. For clarity, we will reference the prior working paper Ref.~\cite{arxiv} multiple times in this chapter, to indicate parts where the two overlap.\\
\\
We now discuss the layout of this chapter -- which is particularly important to establish because our goal is to make it accessible to the broadest possible readership without compromising the details necessary to prove our claims. To achieve this, we have sectioned the presentation accordingly. Specifically, we dedicate Sec.~\ref{Sec_2} to a general overview of the field and our results. This can be read as a self-contained chapter by a general audience, with the intention that it explains what we did to answer this question, how we did it, what we found, and what it means -- but without requiring too much technical detail. Section~\ref{Sec_3} then re-analyzes the problem at a deeper level in terms of the underlying physics. It intends to provide a specialist discussion of the physics of generating entangled states in light-matter systems, and hence the significance of the results that we present. Hence we have aimed~\ref{Sec_3} at a physics audience who wants to see the nuts and bolts of the problem and the specific Hamiltonian that we solve numerically. Section~\ref{Sec_4} provides a detailed discussion of the results in the language of quantum many-body physics. Section~\ref{Sec_5} provides a statement of our main conclusions.  

\section{Non-specialist overview of our analysis and findings}\label{Sec_2}

This section aims to give a general overview of what we did, why we did it, what results we obtained, and why they are important. We intend it to be readable as a self-contained section without requiring too much technical knowledge. Hence we will be fairly liberal and lax in our use of everyday language and terminology to avoid technical jargon \cite{arxiv}. Any reader requiring more rigorous discussion and details is referred to the later Secs. 3 and 4, which can again be read as their own self-contained (but now highly technical) article.

\begin{figure}[t!]
\begin{center}
\includegraphics[width=1.0\linewidth]{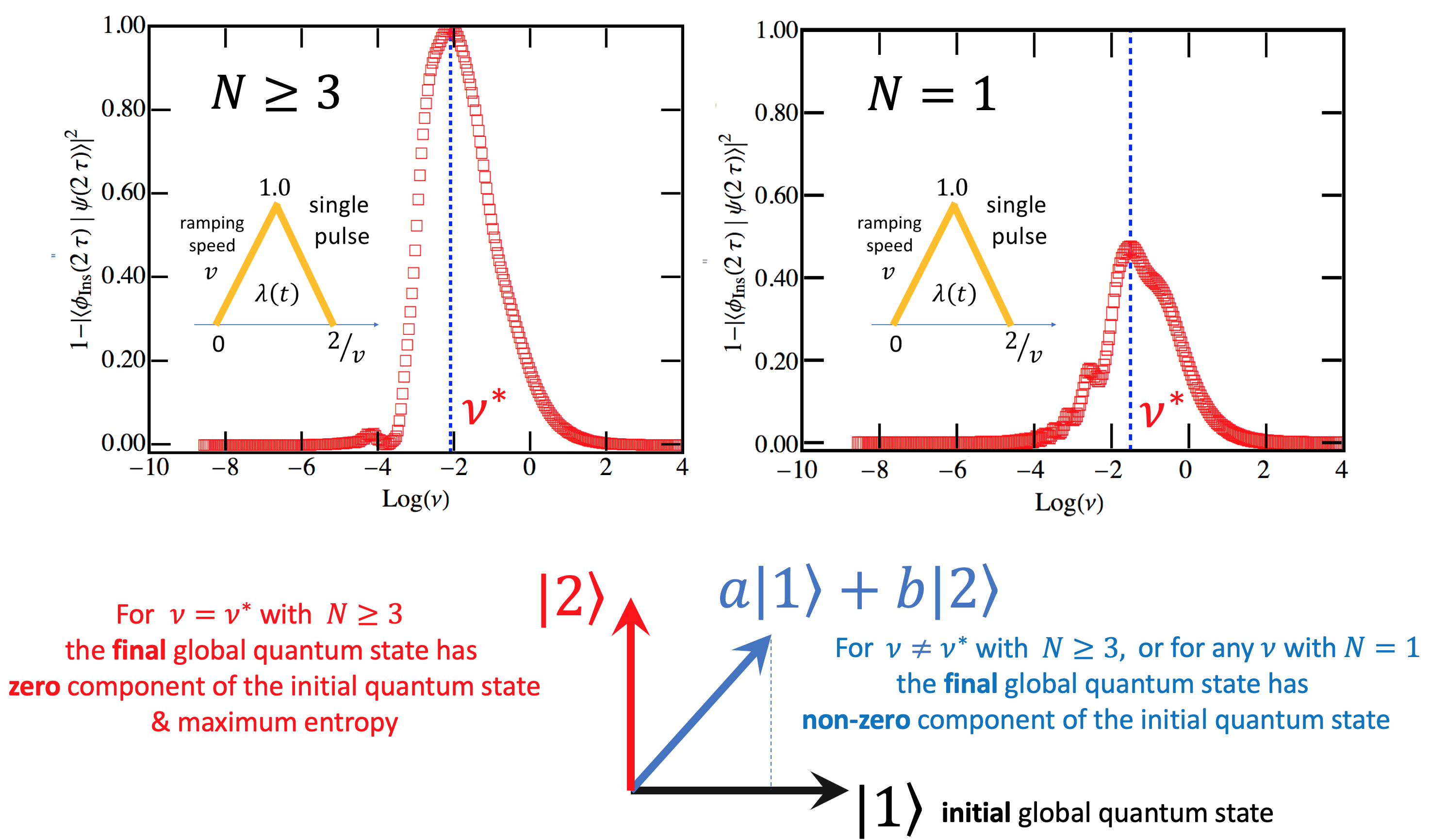}
\end{center}
	\caption{
	{\bf Disruption caused by an adversary.} {\bf Top}: Complete orthogonality to the initial global quantum state (which in practical terms translates to total disruption and hence maximal possible damage~\cite{arxiv})  is produced by a hostile group (left panel) applying a pulse attack $\lambda(t)$ at pulse speed $\upsilon=\upsilon^*$ simultaneously to  $N$ qubits in a bosonic-mode as in Fig.~\ref{fig1}(a). Since the final global quantum state for $\upsilon=\upsilon^*$ has no overlap with the initial one, the initial state cannot be filtered out from the final one -- hence maximum disruption. This is fundamentally different from an attack by a lone `Eve' (i.e. single adversary and hence $N=1$) for whom there is no $\upsilon$ that produces the same effect (right panel). The curve in the left panel is for group size $N=3$ but is visually the same for any $N\geq 3$. We use an initial global quantum state of no photons here for simplicity, but our results can be generalized. {\bf Bottom}: Schematic representation of this disruption.}
 	\label{fig2}
\end{figure}

\begin{figure}[t!]
\begin{center}
\includegraphics[width=1.0\linewidth]{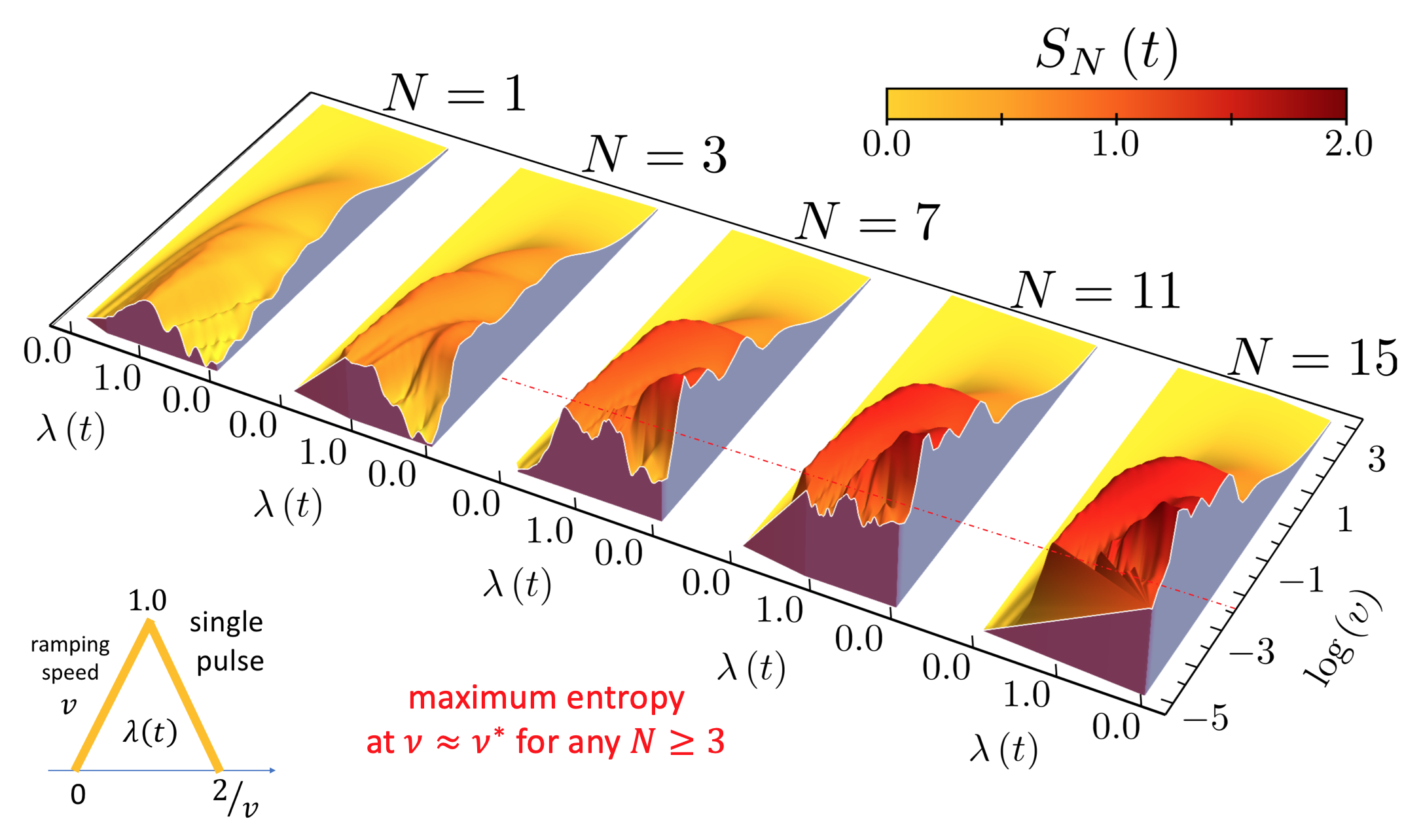}
\end{center}
	\caption{\label{fig3}{\bf Disorder (i.e. entropy) induced by attack}. Von Neumann entropy $S_N$, which is a measure of the disorder in the quantum state, shown throughout the attack (i.e. as a function of $\lambda(t)$ and hence starting and ending at $\lambda(t)=0$) for a given pulse speed $\upsilon$ and a given number of qubits $N$. The number of attackers (adversaries) is the same as the number of qubits since each adversary chooses their own qubit. The general case $N\geq 3$ leads to a final global quantum state with maximum entropy when $\upsilon=\upsilon^*$ within numerical error, and hence pulse duration $2/\upsilon^*$. Our numerical calculations suggest that essentially the same result holds for all higher $N$, with the additional impact that the value of the maximum entropy (and hence the disruption) increases with $N$ (i.e. peak becomes increasingly red in the figure).} 
\end{figure}

\begin{figure}[t!]
\begin{center}
\includegraphics[width=0.9\linewidth]{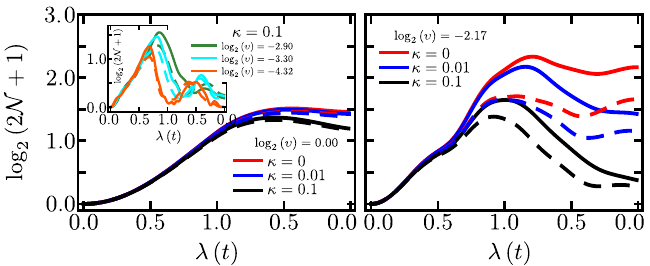}
\end{center}
	\caption{\label{fig4} {\bf Robustness to decoherence}. The quantum logarithmic negativity (which can loosely be thought of as reflecting the amount of quantum entanglement) for an open version of our system (vertical axis) as a function of time during the pulse attack, and hence as a function of $\lambda(t)$. The quantum logarithmic negativity has become a widely accepted entanglement measure for an open system and incorporates the effects of natural decoherence and losses from the cavity. Results are shown for adversarial attacks on systems of $N=5$ (solid lines) and $N=11$ (dashed lines) qubits and for several values of decoherence $\kappa$. The results show a surprising robustness against decoherence and losses that increase with the number of qubits $N$ in the quantum information processing system or quantum computer.}
\end{figure}

We now explain why we focus on adversaries varying the qubit-bosonic-mode interaction $\lambda$ as their method of attack~\cite{arxiv}. Much of the existing analysis of quantum attacks and security is done based on the assumption that so-called gates exist (i.e. specific logic operations are possible). But such gates have to be achieved through some kind of interaction. This means that any quantum adversarial group will need to introduce a term into the system's equation (i.e. Hamiltonian) which will, they might hope, operate in a way similar to some set of gate operations. $\lambda(t)$ plays this role in Fig.~\ref{fig1}(a) and Eq.~\eqref{hdic}. If they are looking for large disruption, the question that we address in this paper is then how bad this could be. It is basic science that perturbing (e.g. attacking) a system involves interacting with it, hence our approach to analyzing the behavior of the quantum system with the interaction included makes sense. Second, it is basic math that it is impossible to recover the unique original quantum state if the final quantum state is orthogonal to it. Just imagine 2 vectors chosen from M orthogonal vectors. Picking one (the final state) says nothing about the other (original state). It is impossible to deduce the second from the first. 
The major surprise from our work is that such an orthogonal state, and hence maximal disruption, can emerge from such adversarial attacks using a simple pulse interaction in qubit-bosonic-mode systems with $N\geq 3$ qubits (see Fig.~\ref{fig2}). This finding is only made possible because of our detailed numerical calculations of the quantum many-body out-of-equilibrium system.\\ 
\\
Figure~\ref{fig1}(a) shows arguably the ultimate limit of such quantum technology in which an extended geographical space serves as a quantum cloud within which the quantum state is kept coherent: specifically, a spatially extended cavity with bosonic modes that contains an arbitrary number $N$ of qubits (two-level systems)~\cite{Ritter,Mirza,Lloyd2}. Given the current experimental success in distributing entangled photons over large distances~\cite{Hofmann72,hensen,SimonPRL}, photons will likely provide the quantum `glue' that binds together large-scale future quantum technologies globally, including a quantum Internet-of-Things. To be fully quantum mechanical, the system size should be within the coherence length of non-local correlations. For recent medium- to long-range practical systems, we refer to Refs.~\cite{hensen,Yin1140,lidar1}. An intermediate version akin to Fig.~\ref{fig1}(b) could be achieved more quickly, in which smaller versions of such quantum node systems are interconnected through separate quantum or classical communication channels. Though still extremely challenging, we note that quantum coherence within cavities has already been demonstrated experimentally in a wide range of laboratory systems including solid-state and quantum optics~\cite{AgarwalPRL1984,HerreraPRL2016,schneider2012rpp}.\\
\\
We now provide a brief description of our underlying theoretical model that provides our conclusions. This model is presented in detail in Sec.~\ref{Sec_3}, while its output is discussed in more detail in Sec.~\ref{Sec_4}. It is purposely chosen to be simple enough that it allows the development of quantitative results and intuition, and yet is realistic enough to capture the highly non-trivial empirical nature of quantum light-matter interactions~\cite{DickePR1954,Klauss,Lieb1973}. Since the $N$ `Eve' adversaries are ultimately humans or machines, we allow each of them to choose a single qubit from the $N$ available: they are each going to then manipulate the interaction between their chosen qubit and the bosonic mode using a pulse-shaped  time-dependent interaction form as in Fig.~\ref{fig1}(a). Strictly speaking, our results are unchanged irrespective of whether each of these interactions is controlled by a single human, a set of humans, another machine, an algorithm, or a bot. We assume that each individual has modest capabilities: they can communicate classically once at some stage before the attack, concerning when to interact their qubit with the bosonic mode and with what temporal profile $\lambda(t)$. This classical communication can be achieved in many mundane ways, e.g. diffusion, announcement, or pass-it-on meaning that the $N$ adversaries do not need to know each other or show any active coordination or collaboration during the subsequent pulse attack when $\lambda(t)$ is non-zero. The interaction pulse $\lambda(t)$ can have a very short duration, e.g. a fraction of a second is possible using current cavity technology~\cite{XiuPRA2016,GuerinPRL2016,KlinderPNC2015,WillPRL2016,ExpDicke4}. Our results are similar for all up-down profile shapes and so we assume a triangular one for simplicity. We assume the resonant condition $\epsilon=\omega$ which has indeed been demonstrated in many laboratory systems experimentally~\cite{XiuPRA2016,GuerinPRL2016,KlinderPNC2015,WillPRL2016,ExpDicke4}. Though we use a specific form of the system equation (i.e. Hamiltonian), similar overall conclusions should follow from many variants due to an established universal dynamical scaling \cite{Acevedo2014PRL,Gomez_Ruiz_2023, Gomez_Ruiz_2018, Gomez_Ruiz_2017, Gomez_Ruiz_2016} and the fact that they typically generate similar types of phase diagrams, and hence have similar collective states in the static $\lambda$ limit \cite{a,b,c,d}. Indeed, we have already shown elsewhere in the science literature that there is a universal dynamical scaling behavior for a particular class concerning their near-adiabatic behavior, in particular the Transverse-Field Ising model, the Dicke Model and the Lipkin-Meshkov-Glick model~\cite{Acevedo2014PRL}.\\ 
\\
The disruption inflicted on the global quantum state by a group of $N=3$ adversaries (since there are $N=3$ qubits), utilizing a pulse interaction $\lambda(t)$, is vividly depicted in Figure~\ref{fig2} (top left panel). This scenario stands in stark contrast to the outcome observed when dealing with a lone wolf attacker ($N=1$ qubit), as illustrated in the top right panel of Figure~\ref{fig2}. Remarkably, no conceivable choice of $\upsilon$ for $N=1$ can reproduce the disruptive effect caused by the collective action of $N=3$ adversaries, nor can any number of asynchronous lone wolf attacks achieve a comparable outcome.\\
\\
These findings hold true not only within the system  depicted in Figure~\ref{fig1}(a) but also within each individual quantum cavity delineated in Figure~\ref{fig1}(b). The consequential disorder, often quantified as ``entropy", induced into the system is visually represented in Figure~\ref{fig3}.\\
\\
An intriguing observation emerges when examining the relationship between the pulse speed $\upsilon$ and the manifestation of unique orthogonality for $N\geq 3$, as depicted in the left panel of Figure~\ref{fig2}. It aligns remarkably closely, within numerical error, with the value at which the induced entropy reaches its maximum, as evidenced in Figure 3. This alignment provides compelling support for our central conclusion: groups comprising $N\geq 3$ adversaries pose a distinct and formidable disruptive threat to the integrity of future quantum systems.\\
\\
Our main findings can be summarized as follows. First, adversarial `Eve' groups employing intermediate attack speeds $\upsilon^*$ will be able to generate maximal global quantum state disruption in systems containing $N\geq 3$ qubits -- which means in practice {\em all} future quantum technology systems since none are likely to only have 1 or 2 qubits alone. Specifically, the final global quantum state contains zero component of the initial one and the corresponding Von Neumann entropy (which is indicative of the amount of disorder generated in the initial quantum state) is also a maximum within numerical uncertainty (Figs.~\ref{fig2} and~\ref{fig3}). No amount of filtering can then recover the initial global quantum state. Second, within numerical uncertainty, the speed at which this maximal disruption is obtained $\upsilon^*$ is insensitive to $N$ for $N\geq 3$, meaning that a general strategy exists that attackers can easily copy (i.e. plug-n-play) without needing to generalize or understand the underlying many-body quantum mechanics. Third, the attack requires no real-time communication, collaboration, or cooperation between any of the  adversaries. They just need to agree ahead of time when to start the pulse, its maximum value $\lambda=1$, and the speed $\upsilon$ and hence duration $2/\upsilon$  of the pulse. From then on, they can operate independently as clock-synchronized individuals from any geographical location within one of the cavities in Fig.~\ref{fig1}.\\
\\
Fourth, this attack leaves no clues that it has occurred based on the Hamiltonian (i.e. the Hamiltonian that we show later in Eq.~\eqref{hdic} is unchanged after the pulse since $\lambda(t)=0$ before and after) or based on the purity of the corrupted quantum state  (i.e. quantum state purity remains one as long as there is no decoherence) and hence the attack can be practically impossible to trace. Fifth, the attack can be over within a fraction of a second since interactions between material qubits and boson cavity modes can be switched on and off very quickly using current technology~\cite{KlinderPNC2015,WillPRL2016,ExpDicke4}. Sixth, the impact of the attack (Fig.~\ref{fig4}) as measured in terms of an open-system generalization of the entropy from Fig.~\ref{fig3}, appears to become more robust to natural decoherence as the size of the quantum information processing system or quantum computer $N$ increases further beyond 3 to higher numbers (i.e. higher $N$).\\ 
\\
In stark contrast, lone `Eve' attackers with no synchronization of their pulses, will achieve a maximum disruption that is not only smaller (Fig.~\ref{fig2}, top right panel) but which is fundamentally different in that the final global quantum state at $\upsilon=\upsilon^*$ still contains a finite amplitude of the initial state, and hence the correct state (i.e. initial state) can in principle be recovered using purification or distillation schemes. The perfect disruption produced by a synchronized attack on $N\geq 3$ qubits could not strictly be achieved by any number of lone attackers, even if they each attack many times since the overlap (Fig.~\ref{fig2}, top right panel) will always be non-zero. If we were to empower a single lone attacker with an extraordinary potential impact through a much larger $\lambda$, then it is in principle possible -- but for $\lambda=1$ we have shown that there is a fundamental difference between the two scenarios. This qualitative difference is also in complete contrast to classical conflict theory~\cite{Lanchester} where the disruption caused by adversarial  groups of size $N$ is assumed to simply scale as $N$, $N^2$ or at most  $N^\delta$ ($\delta>2$) for any $N$.\\ 
\\
Our final finding concerns the scenario in which the $N$ qubits happen to be clustered together in some way, and hence the $N$ adversaries that operate them will also be clustered together in the same way during the pulse attack. It turns out that such clustering of adversaries is actually a preferred pattern for operation of groups of adversaries -- and remarkably, so too it has been shown experimentally that such clustering of qubits can give an enhanced coherence in the many-body quantum state -- hence this coincidence could enhance the impact of such attacks. Specifically, previous work \cite{Spagat,us4,us5,us3,Neil2}  has shown that modern adversarial groups tend to show a common power-law-like cluster distribution with exponent near $2.5$ both offline and online \cite{Spagat,us4,us5,us3,Neil2}, and hence that this is a natural pattern for how humans self-organize for adversarial activities. Such clustering might at first be expected to degrade the post-attack quantum state, akin to how islands of impurities might be expected to enhance single-particle scattering, and hence weaken the finding that the final state post-attack is orthogonal to the initial one. However, it has been shown experimentally \cite{Fratini} that the coherence of a global many-body quantum state is actually {\em strengthened} if there are matter clusters with an approximate 2.5 power-law size distribution \cite{Fratini,Yossi}. Taking the $N$ adversaries as similarly clustered and allowing these clusters to generate a constant low-level background interaction with the quantum cloud -- like the impurity clusters in the experiment in Ref. \cite{Fratini} -- then the coherence and hence orthogonality of the final state should be similarly {\em strengthened} by any such adversarial clustering.\\ 
\\
Countering this threat properly will require a new understanding of time-dependent quantum correlations in many-body light-matter systems. In the meantime, we note that the technology in Fig.~\ref{fig1}(b) will likely arrive before that in Fig.~\ref{fig1}(a), meaning that a network architecture of connected cavities will be built first. In such a scenario, it would be possible to build redundancy into the network: then even if a perfectly orthogonal final state is generated within one cavity by a adversarial group attack at speed $\upsilon^*$, it would be unlikely that this would be achieved simultaneously within a separate cavity, and so error correction schemes could be carried out across all cavities where each cavity is now a network node, and each node is crudely treated as a two-level system.
\section{Detailed physics behind our calculations}\label{Sec_3}

This section analyzes in depth the calculations that we perform and the results that we obtain. It can be read as a stand-alone technical physics report, together with the detailed results in Sec.~\ref{Sec_4}.\\ 
\\
We first present a deeper discussion of the relevance of light-matter systems as our focus of attention. In the implementation of quantum communication infrastructures, diverse strategies have been employed. One approach involves setting up ad-hoc dark fiber networks designed solely for quantum communications. Another method integrates cutting-edge quantum technologies, such as Quantum Key Distribution (QKD) devices and fiber optics-based quantum links, into existing operational communication networks. This integration allows for the coexistence of quantum and classical links within the same infrastructure. The utilization of ad-hoc dark fiber networks ensures the exclusive use of these dedicated channels for quantum communications, minimizing interference and enhancing the overall security of quantum information transfer. On the other hand, the integration of quantum technologies into existing communication networks represents a more versatile approach, capitalizing on the infrastructure already in place. This strategy facilitates a seamless incorporation of quantum capabilities while sharing resources with traditional communication links. By exploring these diverse strategies, the implementation of quantum communication infrastructures can be tailored to specific needs, balancing considerations of security, efficiency, and adaptability within the broader landscape of communication networks.\\
\\
Light has long been regarded as the most crucial component of information transmission, particularly since the advent of relativity. Photons may travel practically inconceivable distances, and they rarely interact among themselves unless matter particles in different environments mediate that interaction. Particularly, quantum information processing requires significant matter-light interactions. The idea of implementing selected photon interactions in the quantum domain has attracted a lot of attention lately. In terms of interacting photons, the development of active quantum matter seems reasonable involving crystalline states of light or even topological phases.

\subsection{Temporal coupling $\lambda(t)$ from driving field}\label{Sec_3_1}
Now we discuss the light-matter system and hence justify our use of a time-dependent $\lambda(t)$ in our model Hamiltonian (see later Eq.~\eqref{hdic}) in order to represent the pulse-driven light-matter system -- and hence why $\lambda(t)$ is the weapon used by each of the adversaries on their chosen qubit ($N$ qubits total). For quantum systems embedded in complex environments, where extra degrees of freedom modulate
the interaction between the quantum system of interest and a large reservoir, effective non-Markovian behaviors in the quantum system dynamics arise even though the reservoir itself can be described within a Markovian approximation.  Consequently, although the phase imprinted by the excitation laser is indeed lost during the first steps of electron-exciton relaxation from the high-energy sector, this is not a sufficient reason to exclude any coherent-like behavior in the relaxing dynamics. Indeed it can be shown that for a wide class of phase-mixed states of the pump modes, results for the signal population can be obtained that are identical to those for a coherent population of those modes. To clarify this critical point, we now show that our basic premise is justified for a variety of reasons. According to the extensive literature concerning light-matter Hamiltonians, in the classical limit, the system can be considered equivalent to two coupled harmonic oscillators. This information is enough to gain analytical insight into the solution of the resulting quadratic system. The driven system in this limit is described by two coupled harmonic oscillators with a time-dependent coupling frequency. Consequently, for this purpose, we will consider a simplified model for the parametric process that contains just $3$ boson modes (for the sake of simplicity we describe now the $N$ dimer subsystem in the low excitation limit as an effective boson $b$ mode), as described by the Hamiltonian:
{\small\begin{equation}
\hat{H}=\omega_a \hat{a}^{\dag}\hat{a}+\chi \left(\hat{a}^{\dag}\hat{a}\right)^2+\omega_b \hat{b}^{\dag}\hat{b}+\omega_c \hat{c}^{\dag}\hat{c}+g\left(\hat{a}^{\dag}\hat{b}^{\dag}\hat{c}^2+\hat{a}\hat{b}\hat{c}^{\dag 2}\right) 
\end{equation}}
where the operators $\hat{a}$, $\hat{b}$ and $\hat{c}$ correspond to bosonic modes (e.g. vibronic, exciton, and high energy exciton modes) where we allow for anharmonic terms of strength $\chi$.
We now consider the effect of the pump state on the dynamics of this simple, but representative, model. In particular, we consider the excitation of high-energy electron states, which indirectly feeds (through the relaxation process) an effective pump reservoir that follows the applied radiation pulse shape.  We assume that $\left(\hat{a}^{\dag}\hat{b}^{\dag}\hat{c}^2+\hat{a}\hat{b}\hat{c}^{\dag 2}\right) = h(t)(\hat{a}^{\dag}\hat{b}^{\dag}+\hat{a}\hat{b})$ where $h(t)$ represents the applied pulse shape. It is usually argued that the expectation value $\langle \hat{c}^{\dag 2}\rangle$ ($\langle\hat{c}^{2}\rangle$) is different from zero only if the high energy reservoir states have coherent populations. Since the laser pulse excites electrons at a higher energy, the excess energy might be expected to relax into the exciton region giving rise to a coherent interaction. However this is not necessarily the case: after non-resonant excitation, the phase imprinted by the excitation laser is generally lost. The appearance of a well-defined phase is often regarded as the true characteristic feature of a coherent state. However, a careful analysis of unitary dynamics from mixed states, such as those produced by incoherent relaxation processes, shows that coherent-like behaviors can often be obtained. To justify this last claim we compute the time evolution of observables under two kinds of pump initial states: (i) A pure initial state like $\left|\right.\Psi\rangle=\left|\right.0_a\rangle \left|\right.0_b\rangle \left|\right.\alpha_c\rangle$, denoting the vacuum state for both modes and a pump coherent state. (ii) A statistical mixed state with no phase information at all, given by a density matrix $\hat{\rho}_P=\int_0^{2\pi}d\theta P(\theta)\hat{\Pi}_p(\theta)\left|\right.\Psi\rangle\langle\Psi\left|\right.\hat{\Pi}_p^{-1}(\theta)$, where $\hat{\Pi}_p(\theta)=e^{i\hat{N}\theta}$, with $\hat{N}=\hat{a}^{\dag}\hat{a}+\hat{b}^{\dag}\hat{b}+\hat{c}^{\dag}\hat{c}$, denotes a phase smearing operator, given the fact that it takes a pump coherent state $\left|\right.\alpha_c\rangle$ to a different phase coherent state $\left|\right. e^{i\theta}\alpha_c\rangle$. The function $P(\theta)$ fixes the pump phase smearing effect with $P(\theta)\geq 0$ and $\int_0^{2\pi}d\theta P(\theta)=1$. Since $\left[\hat{H},\hat{N}\right]=0$, it follows that the time-evolution operator $\hat{U}(t)=e^{-i\hat{H} t}$ commutes with the phase smearing operator $\hat{\Pi}_p(\theta)$. It is now an easy task to obtain for any  observable like $\hat{a}^{\dag k}\hat{a}^{l}$, the time-evolution as
{\small\begin{equation}
\begin{split}\label{Eq4}
\langle\hat{a}^{\dag k}\hat{a}^{l}\rangle_P=& Tr\{\hat{a}^{\dag k}\hat{a}^{l} \hat{\rho}_P(t) \}\\
=&\int_0^{2\pi}d\theta P(\theta)Tr\{\hat{U}^{-1}(t)\hat{\Pi}_p^{-1}(\theta)\hat{a}^{\dag
k}\hat{a}^{l} \hat{\Pi}_p(\theta)\hat{U}(t)|\Psi\rangle\langle\Psi|\}\ .
\end{split}
\end{equation}}
Since $ \hat{\Pi}_p^{-1}(\theta)\hat{a}^{\dag k}\hat{\Pi}_p(\theta)=e^{-ik\theta}\hat{a}^{\dag k}$ and $ \hat{\Pi}_p^{-1}(\theta)\hat{a}^{l}\hat{\Pi}_p(\theta)=e^{il\theta}\hat{a}^{l}$ it follows that
{\small
\begin{equation}\label{E5}
\langle\hat{a}^{\dag k}\hat{a}^{l}\rangle_P=\langle\hat{a}^{\dag k}\hat{a}^{l}\rangle_0\int_0^{2\pi}d\theta
P(\theta)e^{-i(k-l)\theta}
\end{equation}}

\noindent where $\langle\hat{a}^{\dag k}\hat{a}^{l}\rangle_0$ corresponds to the initial state with the pump in a coherent state. It is evident that the population dynamics of the subsystem ($k=l=1$) is fully insensitive to this class of phase smearing in the pump state, $\langle\hat{a}^{\dag}\hat{a}\rangle_P=\langle\hat{a}^{\dag}\hat{a}\rangle_0$, as well as other vibrational correlations as long as the pump phase smearing probability $P(\theta)$ remains practically constant. The main physical ingredients of a general, complex light-matter system can therefore be captured by this simple $3$-mode Hamiltonian. Therefore we can conclude that for a wide class of coherent pump-plus-relaxation process conditions, our main results are indeed meaningful. Hence the replacement of $\hat{c}$-pump operators by complex numbers -- which consequently yields a time-dependent coupling strength $\lambda(t)$ -- is justified. Also, the range of validity of our assumption is the same as the usual one for the parametric approximation which requires a highly populated coherent state, $|\alpha_c|\gg 1$, and short times, $gt\ll 1$. These conditions are precisely identical to those under which our model fits with previous studies of coherence generation: high excitation and rapid relaxation dynamics. Therefore, there is indeed a formal justification for reducing the last terms to $g h(t)(\hat{a}^{\dag}\hat{b}^{\dag}+\hat{a}\hat{b})$ where $h(t)$ represents the applied pulse shape -- hence justifying the time-dependent interaction $\lambda(t)\sim g h(t)$.\\
\\
Next, we discuss the fact that any incident electromagnetic (light) field ${\vec E}$ generates an internal polarization field ${\vec P}$ within the material, given exactly by Maxwell's Equations. The equation describing the time-domain behavior in a general, anisotropic and nonlinear medium subject to a general time and position-dependent light field $\vec E$ is given by:
\begin{equation}
 \nabla \times \nabla \times \vec E + \mu_0\sigma\frac{\partial \vec E}{\partial t} +\mu_0\frac{\partial^2 {\vec{\vec\epsilon}}\cdot\vec E}{\partial^2 t}=-\mu_0 
\frac{\partial^2 {\vec P}}{\partial^2 t}
\end{equation}
in which the standard symbols have their well-known meaning from electromagnetic theory (e.g. $\vec{\vec\epsilon}$ is a complex second-order tensor). If the medium is lossless then $\sigma=0$ and so this equation can be rewritten as:
\begin{equation}
[ \nabla \times (\nabla \times)  + \frac{1}{\epsilon_0 c^2} 
\frac{\partial^2}{\partial t^2} {\vec{\vec\epsilon}} \cdot ] {\vec E}=-\frac{1}{\epsilon_0 c^2}\frac{\partial^2 {\vec P}}{\partial^2 t} \ .
\end{equation}
Though nonlinear and anisotropic in general, the presence of $\partial^2/{\partial t^2}$ terms for ${\vec E}$ and ${\vec P}$ in both equations means that a pulse in ${\vec E}$ will generate a similar pulse in ${\vec P}$, and hence a pulse in the internal electric field dynamics coupling the electronic and vibrational systems (i.e. a pulse in $\lambda(t)$).\\
\\
Our focus  is on near resonant conditions since these are the most favorable for generating large coherences. We only consider one such resonance for simplicity, however, this can be generalized by matching up different  excitation energies $\epsilon'$, $\epsilon''$, etc. to the nearest vibrational energies $\omega'$, $\omega''$ etc., and then solving the Hamiltonian in the same way for each subset $(\epsilon',\omega')$ etc. For example, if the $N$ components are partitioned into $n$ subpopulations, where each subpopulation has the same resonant energy and vibrational mode but where these values differ between subpopulations, the total Hamiltonian will approximately  decouple into $H^{(1)}\oplus H^{(2)}\oplus H^{(3)}\dots \oplus H^{(N)}$. Any residual coupling between these subpopulations can then be treated as noise.

\subsection{Details of our Hamiltonian model}\label{Sec_3_2}

\begin{figure}[t!]
\begin{center}
\includegraphics[width=0.95\linewidth]{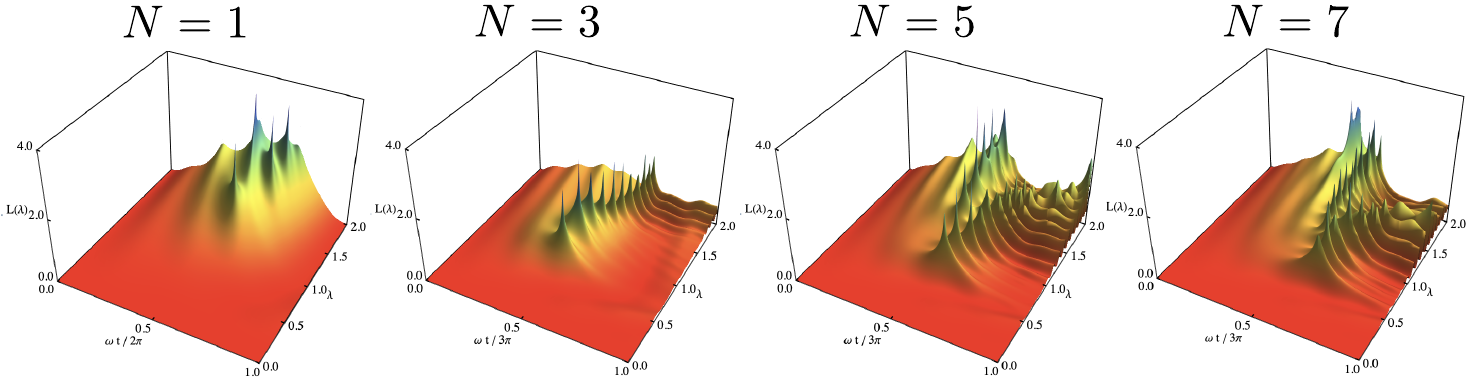}
\end{center}
	\caption{\label{fig5} {\bf Loschmidt Echo (LE) for a constant pulse $\lambda$}. In the realm of deep strong coupling, a sequence of singularity peaks in the Loschmidt Echo becomes apparent, with their emergence dictated by both the coupling parameter $\lambda$ and the time $t$ following the onset of coupling.}
\end{figure}

This section provides a detailed discussion of our model using the specialist language of many-body physics and quantum information. \\
\\
Quantum computers contain approximate two-level systems, where the difference between the two levels is the excitation energy, that are invariably coupled to other bosonic subsystems (e.g. light) as in Fig.~\ref{fig1}. Each adversary can perturb one such qubit system. We describe this using a generalized Dicke Model. Our model features a set of $N$ identical two-level systems (commonly referred to as qubits) each of which is coupled to a single bosonic mode. It can be described by the following microscopic Hamiltonian:

\begin{equation}\label{hdic}
\hat{H}(t)=\frac{\epsilon}{2}\sum_{i=1}^{N}\hat{\sigma}_{z}^{i} + \omega \hat{a}^{\dagger}\hat{a} +\frac{\lambda(t)}{\sqrt{N}}\left(\hat{a}^{\dagger}+\hat{a}\right)\sum_{i=1}^{N}\hat{\sigma}_{x}^{i}\:.
\end{equation}
We purposely avoid common approximations such as the rotating-wave approximation. Here $\sigma_{\alpha}^{i}$ denotes the Pauli operators for qubit $i$ $\left(\alpha=x,z\right)$;  $\hat{a}^{\dagger} \pap{\hat{a}}$ is the creation (annihilation) operator of the radiation field. $\epsilon$ and $\omega$ represent the qubit and field transition frequencies respectively; and $\lambda(t)$ represents the strength of the radiation-matter interaction (i.e. it represents the adversaries' manipulation) at time $t$ which can be varied over time. In many situations such as those we consider here in our work, the dynamics of the Dicke model do not require the consideration of the entire $2^{\otimes N}\otimes \mathbb{N}$ dimensional Hamiltonian. Instead, $SU(2)$ collective operators $\hat{J}_{\alpha}=\frac{1}{2}\sum_{i=1}^{N}\hat{\sigma}^{i}_{\alpha}$ can be used. The Hamiltonian can then be written in the following form:
\begin{equation}\label{Hdicke}
\hat{H}(t)=\epsilon \hat{J}_{z} + \omega \hat{a}^{\dagger}\hat{a} +\frac{2\lambda(t)}{\sqrt{N}}\hat{J}_{x}\left(\hat{a}^{\dagger}+\hat{a}\right)\ \ .
\end{equation}
The static properties of the Dicke Model have been widely studied and characterized in the last two decades~\cite{Nagy2010prl,Das2016njp}. It is well known that, in the thermodynamic limit $N\to\infty$, it exhibits a second-order quantum phase transition (QPT)~\cite{HioePRA} at $\lambda_{c}=\sqrt{\epsilon\omega}/2$ with order parameter $\hat{a}^{\dagger}\hat{a}/J$, separating the normal phase at $\lambda_{c}<\sqrt{\epsilon\omega}/2$ from the superradiant phase in which there is a finite value of the macroscopic order parameter, e.g. finite boson expectation number \cite{Acevedo2014PRL}. When the static coupling parameter is above this critical value $\lambda_c$, the ground state of the system is characterized by a non-zero expectation value of the excitation operators,   
\begin{equation}
\bigl\langle \hat{N}_{b} \bigr\rangle\equiv \bigl\langle \hat{a}^{\dagger}\hat{a}\bigr\rangle \qquad \text{and} \qquad \bigl\langle \hat{N}_{q} \bigr\rangle\equiv \biggl \langle \hat{J}_{z} -\frac{N}{2}\biggr\rangle \:.
\end{equation}
When $\lambda < \lambda_{c}$, the order parameters are zero. Because of this, the region when $\lambda > \lambda_{c}$ is called the ordered or superradiant phase, while the region when $\lambda < \lambda_{c}$ is called the normal phase. Near the phase-boundary in the vicinity of this superradiant phase, there is a dependence of the order parameter as follows: $\bigl\langle \hat{N}_{b} \bigr\rangle\propto\left(\lambda -\lambda_{c}\right)$ and  $\bigl\langle \hat{N}_{q} \bigr\rangle\propto\left(\lambda -\lambda_{c}\right)^{1/2}$~\cite{Emary_PRE}. This power-law behavior is typical of second-order phase transitions where the critical exponents are characteristic of the universality class to which the model belongs. In the ordered quantum phase $\left(\lambda >\lambda_c\right)$, the $\mathbb{Z}_{2}$ symmetry related to parity is spontaneously broken, which originates from the fact that the thermodynamic limit ground state is two-fold degenerate corresponding to the two different eigenvalues of $\hat{P}$. Also at the QPT, the Dicke model presents an infinitely degenerate vanishing energy gap~\cite{Emary_PRE}. \\
\\
We use the Loschmidt Amplitude (LA), which is the overlap between the system's initial quantum state and its time-evolved state, as a new way to quantify quantum vulnerability: when the overlap is zero, it means orthogonality and hence maximum disruption has been created by the adversarial perturbation $\lambda(t)$. The logarithm of this LA value then presents a singularity. We note that the LA is one of the most significant quantities in modern quantum physics research. In particular, it has been used to detect time evolution singularities in quenched qubit-cavity system~\cite{Betancourt_PRA} and even for topological qubits inserted in photon cavities~\cite{Mendez_PRB}. Results for the special case of rectangular pulses are depicted in Fig. \ref{fig5}. Specifically, the LA defined as  ${\cal L}(t)=\langle\Psi(0)|\Psi(t)\rangle$ represents the fidelity of the time-evolved agent state $|\Psi(t)\rangle$ with respect to the initially prepared state $|\Psi(0)\rangle$. When the logarithm of the LA value is close to zero, by this definition, there is a strong overlap between the initial state and the time-evolved state. 
Computationally, we perform a brief quench to the final state, assuming that the two-level systems are initially prepared in the ground state. The return probability of the time-evolved state to the initial state is the definition of the Loschmidt Echo (LE) $\mathcal{L}$ in the simplest of all quench protocols. The appearance of zeros in the Loschmidt Echo implies localization transitions, also known as dynamical phase transitions. It is worth noting that the longer time-evolved states of the post-quench Hamiltonian with time-dependent interactions differ significantly from the original state in which matter and light were created.  As a result, the Loschmidt Amplitude (the scalar product of plane-wave and extended time-evolved states) vanishes at key points, indicating dynamical phase transitions. This means that the system's quantum state is now orthogonal and hence the disruption is maximal.
For the special case where the excitation energy  goes to zero, the LE could be obtained analytically as:
\begin{equation}\label{eq:analitycal}
    {\cal{L}}^{N}(t)=\frac{1}{2^{N-1}}
    \sum_{n=0}^{\left [\frac{N}{2}\right]} \binom{n}{N}
    e^{-\frac{\lambda_M^2}{N}(\frac{N}{2}-n)^2\alpha(t)}
\end{equation}
where
\begin{equation}
\alpha(t)=\frac{1}{3\omega_c^4 t^2}\left[9-12 e^{-i\omega_c t}+3 e^{-2i\omega_c t}-2 i\omega_c t (3-(\omega_c t)^2))\right].
\end{equation}
We leave the pursuit and analysis of such analytical forms to future work.
\section{Results}\label{Sec_4}

\begin{figure}[t!]
\begin{center}
\includegraphics[width=1\linewidth]{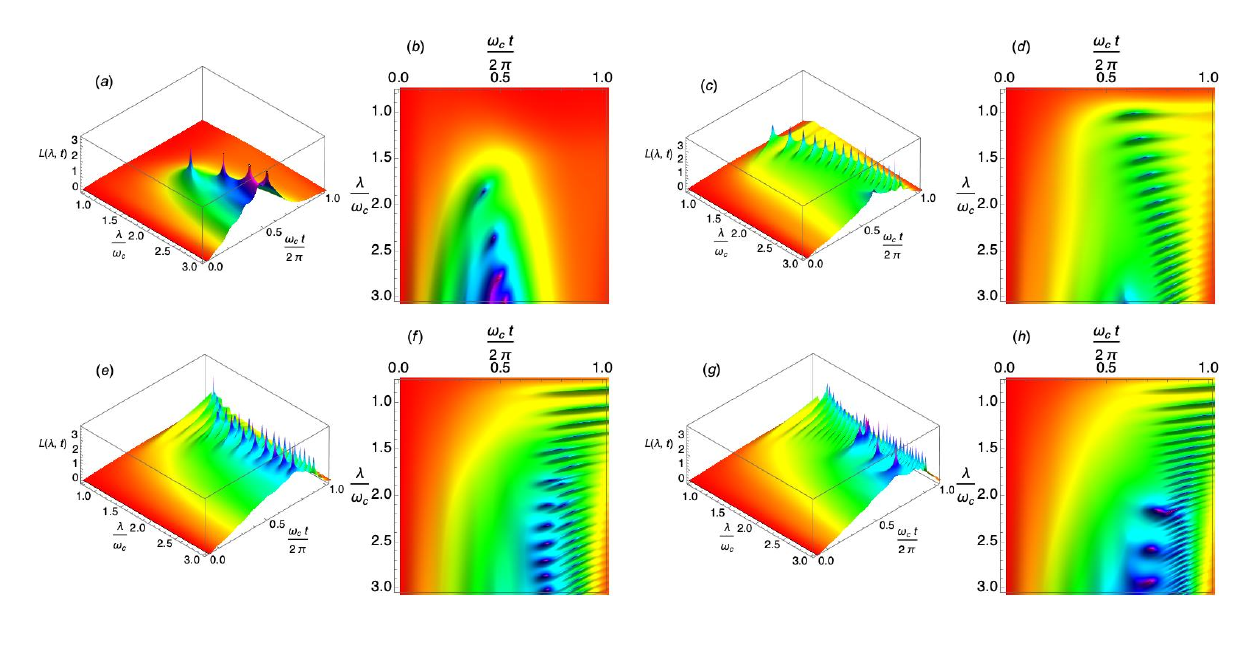}
\end{center}
	\caption{\label{fig6} Detailed output of our model Hamiltonian calculations for $N = 1, 3, 5, 7$. The time-dependent interaction (i.e. adversaries' manipulation) is proportional to $vt$.  For $N\geq 3$, the singularities (and hence the disruption) appear increasingly strong as $N$ increases.}
\end{figure}

We find that the Loschmidt Echo (LE) shows kinks at important points as the system size $N$ grows. At these points, the overlap with the original state is minimal, i.e. the overlap is near zero, hence the logarithm of the overlap shows a feature akin to a finite-size version of a singularity. The dynamics are such that over time, the state will spread in Hilbert space due to the coupling with confined light, reducing the peaks. Crucially, even while the light effectively functions as a dissipative information system, this expansion of cusps is broad and does not require specific adjustments of parameters. 
For a time-independent $\lambda/\omega_c < 1$, the overlap between two unique ground states in the Dicke model approaches zero as the number $N$ of two-level systems rises. Because of this, both of the two ground states are mutually orthogonal in the thermodynamic limit. This effect is known as the Anderson orthogonality catastrophe and occurs in systems with infinitely many degrees of freedom when two physical states that correspond to two arbitrarily near sets of parameters (two arbitrarily comparable physical conditions) become orthogonal states of one another. This effect has previously been investigated in many-body physics. In contrast, a large peak develops in a band where the overlap approaches one for values of $\lambda/\omega_c >1$.  Surprisingly, we find that the system is susceptible to losing the original data for this collection of parameters. Thus, for zeros emerging at larger levels of $\lambda(t)$, we may observe that the composition of the initially prepared state becomes progressively dominated by the contribution of different two-level system states at higher energy states. 
An $N$-qubit circuit with an initial state for the unexcited agent and zero communication photons, with $N = 1, 3, 5, 7, 9$. Fig.~\ref{fig6} shows the LE results, where two unique zones are identified. (1) The LE travels to a single zone (vulnerable zone), and (2) a complicated structure makes the system immune to external perturbations.
The two-dimensional representation of the complicated structure is shown in Fig. \ref{fig6}((b),(d),(f),(h)), on the right side. We have demonstrated that it is possible to achieve a regime  where the light controls the zones in which the system fails to return to the initial state.
This ability to manipulate  the initial state so that it abruptly collapses into orthogonal states as a function of $\lambda$ and $t$, represents a previously unknown threat to quantum technologies.\\ 
\\

Our calculation of the impact of decoherence, as discussed in Sec.~\ref{Sec_2}, uses the density matrix approach of Ref.~\cite{AcevedoPRA2015} and a widely accepted measure of the open-system entanglement, the quantum negativity. 
We find that the presence of decoherence does not change our main conclusions. Not only do our main results survive well with increasing decoherence $\kappa$, the strength and robustness of the many-body coherence both increase with $N$. (For very large $N$, other system-level decoherence mechanisms may of course set in). Different temperatures are simulated by varying the average number of phonons $\bar{n}$, choosing values typical of the low temperatures in most experimental realizations, and including a thermal distribution in the initial density matrix~\cite{AcevedoPRA2015}. Again our main conclusions are qualitatively unchanged.\\ 
\\
Finally, we stress that  our focus here has been on the question of collective coherence in any setting where there is some pulsed perturbation of the system due to adversaries manipulating the system. However the external profile ${\vec E}(t)$ and hence $\lambda(t)$ can be general. It need not be a pulse. Also the application of our system Hamiltonian analysis could be to transport experiments as well as optical experiments, or any combination of these.
\section{Conclusions}\label{Sec_5}
We have addressed a key question for society concerning future quantum technologies: specifically, their vulnerability to collective manipulation. Specifically, we presented and discussed a new form of vulnerability in such systems that we identified based on detailed many-body quantum mechanical calculations. This new vulnerability will enable hostile groups to inflict maximal disruption on the global quantum state in such systems and hence will threaten their core functionality. These attacks will be practically impossible to detect since they introduce no change in the Hamiltonian and no loss of purity; they require no real-time communication; and they can be over within a second. We also predicted that such attacks will be amplified by the statistical character of modern non-state actor groups. A countermeasure could be to embed future quantum technologies within redundant classical networks.\\ 
\\
 We also analyzed more deeply the origins of this disruption by adversaries. Specifically, we showed in Sec. 3 that for a broad range of parameters, the system-level disruption (i.e. transitions) are characterized by curves in the appropriate Loschmidt functions at important points. By examining the numerical solution derived for the Loschmidt Amplitude, we found that the singularities were a component of a more intricate underlying structure in the system's temporal development. We used the Lodschmidt Echo to draw a connection between the macroscopic dynamics and the quantum vulnerability.

\section*{Acknowledgments}
Spanish MCIN supported part of this research with funding from European Union Next Generation EU (PRTRC17.I1) and Consejeria de Educaci\'on from JCyL through the QCAYLE project, as well as MCIN projects PID2020-113406GB-I00 and RED2022-134301-T.  F.J.R. and L.Q. are thankful for financial support from Facultad de Ciencias-UniAndes projects INV-2021-128-2292 and INV-2023-162-2833. N.F.J. is supported solely by U.S. Air Force Office of Scientific Research awards FA9550-20-1-0382 and FA9550-20-1-0383, as well as The John Templeton Foundation. The views and conclusions contained herein are solely those of the authors and do not represent official policies or endorsements by any of the entities named in this book chapter. An initial working paper which forms the basis for part of this paper, was posted as a preprint on the well-known arXiv preprint server: `Quantum Terrorism: Collective Vulnerability of Global Quantum Systems' in 2019 \url{https://arxiv.org/abs/1901.08873}.

\bibliography{Bib_Book_Chapter}{}
\bibliographystyle{apsrev4-2}

\end{document}